\documentclass[12pt]{iopart}
\usepackage{iopams} 
\usepackage{graphicx}
\usepackage{braket}
\usepackage{upgreek}
\usepackage{xcolor}
\usepackage{cite}

\begin{document}

\title[Nanocaivty Enhanced Photon Coherence of Solid-State Quantum Emitters]{Nanocavity Enhanced Photon Coherence of Solid-State Quantum Emitters Operating up to 30 K}

\author[1]{A~J~Brash$^1$, J~Iles-Smith$^2$}
\address{$^1$ Department of Physics and Astronomy, University of Sheffield, Sheffield S3 7RH, UK}
\address{$^2$ Department of Physics and Astronomy, The University of Manchester, Oxford Road, Manchester, M13 9PL, United Kingdom}
\ead{a.brash@sheffield.ac.uk}
\vspace{10pt}
\begin{indented}
\item[]July 2023
\end{indented}

\begin{abstract}
Solid-state emitters such as epitaxial quantum dots have emerged as a leading platform for efficient, on-demand sources of indistinguishable photons, a key resource for many optical quantum technologies. To maximise performance, these sources normally operate at liquid helium temperatures ($\sim 4~\mathrm{K}$), introducing significant size, weight and power requirements that can be impractical for proposed applications. Here we experimentally resolve the two distinct temperature-dependent phonon interactions that degrade indistinguishability, allowing us to demonstrate that coupling to a photonic nanocavity can greatly improve photon coherence at elevated temperatures up to $30~\mathrm{K}$ that are compatible with compact cryocoolers. We derive a polaron model that fully captures the temperature-dependent influence of phonons observed in our experiments, providing predictive power to further increase the indistinguishability and operating temperature of future devices through optimised cavity parameters.
\end{abstract}

%
%
%
%
%

\section{Introduction}

Single, indistinguishable photons are a vital building block for many proposed optical quantum technologies such as optical quantum computing \cite{PhysRevLett.86.5188,walther2005experimental,rudolph2017optimistic}, long range secure quantum networks \cite{munro2012quantum,azuma2015all,10.1002/qute.202100116} and optical quantum metrology \cite{giovannetti2011advances}. Devices based upon epitaxially grown III-V semiconductor quantum dots (QDs) coupled to micro-/nano-photonic structures have emerged as a leading single photon source (SPS), owing to their potential to generate single photons ``on-demand" with high efficiency, purity and indistinguishability \cite{PhysRevLett.116.213601,Somaschi2016,Liu2018,tomm2021bright}. Indistinguishable photons from such sources have facilitated important quantum technology demonstrations, including linear optical quantum computing \cite{PhysRevLett.123.250503,maring2023generalpurpose} and entanglement swapping for quantum communications and networking \cite{PhysRevLett.123.160502,doi:10.1126/sciadv.abe6379}. Furthermore, similar methods can be extended to produce more complex resource states for optical quantum technologies, such as recent demonstrations of entangled graph states \cite{cogan2023deterministic,coste2023high,PhysRevLett.128.233602} where high fidelities are enabled by indistinguishable photons. Beyond QDs, the cavity-emitter concept has also been applied to realise photon sources using quantum emitters in other solid-state hosts such as diamond \cite{benedikter2017cavity}, silicon \cite{redjem2023all} and 2D materials \cite{doi:10.1021/acs.nanolett.2c03151}. At present, III-V QDs offer the most attractive platform due to their large dipole moment and relatively weak phonon coupling at low temperatures, enabling high brightness and indistinguishabilities.

Owing to a desire to minimise potentially detrimental interactions with phonons, studies of indistinguishable photon emission from QD-based SPSs have generally focused on temperatures around $4~\mathrm{K}$ in either open- or closed-cycle helium cryostat systems. Whilst significantly smaller and less complex than the mK dilution refrigerator systems that house superconducting circuits for quantum computing research, these systems still have considerable associated size, weight and power (SWAP) costs. The importance of SWAP requirements becomes particularly clear when contemplating potential usage cases for optical quantum technologies, for instance the tight space and thermal constraints of data centres, or the SWAP-critical environment of satellite communications. An alternative approach is to use a device such as a compact Stirling cryocooler, which are often specified for satellite instruments due to SWAP and maintenance considerations. In a proof-of-concept demonstration with a QD sample, the mean base temperature of such a cryocooler was found to be 28.8 K \cite{10.1063/1.4906548}. Whilst single photon emission has been observed from various types of epitaxial III-V QDs at temperatures reaching as high as $350~\mathrm{K}$ \cite{10.1063/1.1650032,Holmes2016,Kolatschek2021,Laferriere2023}, increased contributions from phonon processes cause a rapid loss of indistinguishability for even small increases above $4~\mathrm{K}$ \cite{PhysRevLett.93.237401,PhysRevLett.116.033601,PhysRevB.97.195432,PhysRevLett.118.233602}.
As such, for future quantum technology applications, a major outstanding challenge is to generate \emph{indistinguishable} photons at temperatures compatible with compact cryocoolers.

\subsection{Real and Virtual Phonon Processes}

\begin{figure}
    \centering
    \includegraphics{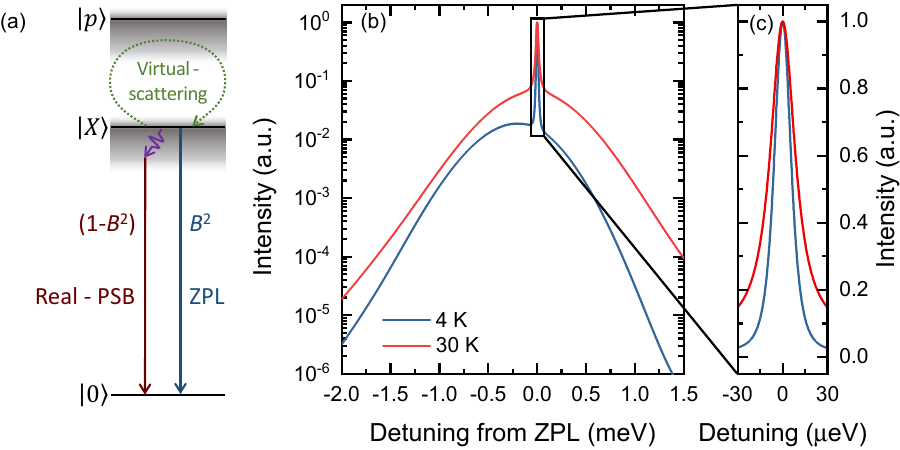}
    \caption{Influence of phonons on the optical transitions of a QD TLS: (a) Energy level diagram of the QD TLS comprising ground $\ket{0}$ and exciton $\ket{X}$ states. Direct decay of the exciton to the ground state results in the familiar zero phonon line with a probability given by the Frank-Condon factor $B^2$. Real transitions corresponding to emission/absorption of a phonon during exciton relaxation lead to emission of a photon with distinguishable frequency, forming a phonon sideband with relative area $(1-B^2)$. Meanwhile, virtual transitions to higher energy states $\ket{p}$ occur through scattering of thermal phonons, broadening the ZPL. (b) Log-linear theoretical spectrum of the QD, showing the narrow ZPL and broad PSB. The PSB area and symmetry both increase noticeably at 30 K (red line) compared to 4 K (blue line). Spectra are produced using the experimental parameters found in this work. (c) Linear-linear close-up of the ZPL, showing thermal broadening at 30 K (red line) compared to 4 K (blue line). The ZPL at 4 K is already significantly radiatively broadened by the inclusion of a Purcell factor of 43.}
    \label{fig:phononspectra}
\end{figure}

For self-assembled III-V semiconductor QDs, the dominant influence of phonons on the spectrum of a QD two level system (TLS) comprising a ground state and the lowest-energy exciton state (s-shell) is electron-phonon coupling through the deformation potential \cite{PhysRevLett.91.127401,PhysRevLett.104.017402}. This interaction occurs between QD-confined electrons and longitudinal acoustic (LA) phonons of the bulk semiconductor material, exhibiting a continuum of phonon states up to a cut-off energy governed by the QD size \cite{PhysRevB.96.245306,doi:10.1080/23746149.2019.1655478,Denning:20}, typically on the order of a few meV \cite{PhysRevLett.104.017402,PhysRevLett.114.137401,PhysRevLett.118.233602,PhysRevLett.123.167403,PhysRevLett.123.167402}. A detailed theoretical treatment of this coupling is described in section \ref{sec:Theory}, however the key result is the emergence of virtual and real phonon-mediated transitions \cite{PhysRevLett.93.237401,Denning:20}, which are illustrated in Fig. \ref{fig:phononspectra}. In the case of the real transitions (linear in phonon operators), the system decays from excited to ground state, with the emitted photon energy (red arrow in Fig. \ref{fig:phononspectra}(a)) reduced or increased by the corresponding emission or absorption of a phonon (purple curly arrow in Fig. \ref{fig:phononspectra}(a)). At $4~\mathrm{K}$, there are very few phonons to absorb and therefore phonon emission processes dominate, giving rise to a broad, asymmetric phonon sideband (PSB). With increasing temperature, both phonon emission and absorption become more probable, but the difference between the two probabilities reduces, leading to a sideband whose area increases and asymmetry decreases with temperature \cite{doi:10.1080/23746149.2019.1655478,Denning:20}, as shown in Fig. \ref{fig:phononspectra}(b). In the absence of a photonic structure, the fraction of light emitted through the PSB is given by $(1-B^2)$, where $B^2$ is termed the Frank-Condon factor.

Meanwhile, virtual processes correspond to virtual transitions (quadratic in phonon operators) between the QD excited state and higher energy electronic states (e.g. p-shell - dashed green arrow in Fig. \ref{fig:phononspectra}(a)) \cite{Denning:20,PhysRevLett.118.233602,PhysRevLett.93.237401}. The effect of these transitions is to produce a temperature-dependent pure dephasing effect, leading to homogeneous broadening of the zero phonon line (ZPL), the well-known Lorentzian spectrum associated with a TLS shown in Fig. \ref{fig:phononspectra}(c).
The width of the ZPL is governed by its coherence time $T_2$:
\begin{equation}
    \frac{1}{T_2} = \frac{1}{2T_1} + \frac{1}{T^*_2},
    \label{eq:T2}
\end{equation}
where $T_1$ is the transition radiative lifetime and $T^*_2$ is the dephasing time associated with the pure dephasing rate. From Eq. \ref{eq:T2} it can be seen that in the absence of any pure dephasing, the coherence time reaches a maximum value $T_2 = 2T_1$, often termed \emph{radiatively limited}. In this limit, photons emitted through the ZPL are perfectly indistinguishable, highlighting the importance of achieving radiatively limited coherence. Since photons emitted into the phonon sideband are completely distinguishable in frequency, the contributions of both types of phonon process can be combined into a general expression for the visibility of two photon interference for photons emitted from a single QD~\cite{Iles-Smith2017}:
\begin{equation}
    V = B^4 \frac{T_2}{2T_1},
    \label{eq:V}
\end{equation}
where $V = 1$ and $V = 0$ correspond to completely indistinguishable and distinguishable photons respectively. A spectral filter whose width and centre frequency matches the ZPL can remove the PSB, increasing to $V = T_2/2T_1$ at the cost of a minimum reduction in efficiency of $(1 - B^2)$ \cite{Iles-Smith2017}.

\subsection{Phonon Processes in Quantum Dots}

The influence of phonon processes on the emission properties of III-V QDs is well studied. Whilst it was established that phonon broadening of the ZPL is essentially negligible at $\sim 4~\mathrm{K}$, it rapidly becomes significant as $T$ increases, leading to a broadening which exceeds the radiative limit by more than a factor of 10 by $50~\mathrm{K}$ \cite{PhysRevLett.93.237401}. However, studies mainly focused on achieving the radiative limit in the low temperature regime where phonon broadening could be neglected, with this ultimately being successful through material quality improvements removing other unwanted environmental effects such as charge noise \cite{Kuhlmann2013}. 
With essentially radiatively limited ZPL emission in the low temperature limit, attention turned to PSB processes as the limit to photon indistinguishability~\cite{Iles-Smith2017,Gustin18}. For InGaAs QDs, a typical value of $B^2$ is around 0.9, limiting unfiltered $V$ to 0.81. To overcome this, QDs were integrated with optical micro-/nano-cavities \cite{PhysRevLett.116.213601,Somaschi2016,Liu2018,tomm2021bright}, where the combination of Purcell enhancement and spectral filtering can remove some of the sideband photons with lower losses than simple spectral filtering \cite{Iles-Smith2017,Gustin18,PhysRevLett.118.253602}. It is important to note however that even with such cavity coupling, there remains a fundamental trade-off between efficiency and indistinguishability, even for ideal cavity parameters \cite{Iles-Smith2017}.

Moving beyond the low temperature limit, several studies have considered the temperature-dependent coherence of photons emitted by QDs in the absence of any significant Purcell enhancement, with all studies observing a rapid decrease in indistinguishability as temperature is increased \cite{PhysRevLett.93.237401,PhysRevLett.116.033601,PhysRevB.97.195432,doi:10.1021/acsphotonics.6b00707}. Theoretical modelling has revealed that both real and virtual phonon processes contribute to this trend \cite{Denning:20,PhysRevLett.118.233602}. A potential strategy to reduce these temperature-dependent effects is again to couple the QD to an optical cavity. In addition to the aforementioned filtering of the PSB photons, for appropriate parameters, the cavity also induces a Purcell enhancement ($F_P$) of the QD emission rate ($F_P/T_1$). From Eq.~\ref{eq:T2}, it can be seen that this enhancement reduces the degredation of the coherence time ($T_2$) for a given pure dephasing rate ($1/T^*_2$), offering the potential to suppress the influence of the virtual phonon transitions. Measurements of a QD-micropillar device with a Purcell factor of 20 exhibited significantly weaker degradation of the emitted photon coherence in the $9$ - $18~\mathrm{K}$ range \cite{PhysRevLett.118.253602}, supporting this prediction. However, the maximum temperature reached in this study ($18~\mathrm{K}$) is still well below the base temperature of a compact cryocooler ($28.8~\mathrm{K}$ \cite{10.1063/1.4906548}), limited by the electrical tuning range required to maintain QD-cavity resonance as the QD redshift grows non-linearly with temperature \cite{PhysRevB.72.085328,Kroner2009}. Furthermore, the maximum Purcell factor attainable in a micropillar cavity is restricted by the increased mode volume compared to nanocavity structures such as photonic crystal cavities (PhCCs) \cite{Somaschi2016,Liu2018}.

In this work, to overcome these limitations and improve the indistinguishability of photons emitted at temperatures compatible with compact cryocoolers, we employ low mode volume H1 PhCCs, fabricated on a QD wafer that achieves $>2~\mathrm{meV}$ QD tuning range using thick AlGaAs tunnelling barriers.
Exploiting these favourable properties, we study the photon coherence of a QD-PhCC device with large Purcell enhancement ($F_P = 43$) over the range 4 - 30 K. By using a novel technique based on time-domain measurement of the first-order correlation function under weak resonant excitation, we simultaneously resolve the real and virtual phonon contributions in a single experiment, unlike previous experiments based on two-photon interference that cannot separate these processes. Owing to the large Purcell enhancement, $T_2/2T_1$ at $25~\mathrm{K}$ is only $7.5~\%$ lower than at $4~\mathrm{K}$, whilst at a temperature of $30~\mathrm{K}$ that is compatible with compact cryocoolers, $T_2/2T_1$ is doubled compared to previous measurements of a QD without an optical cavity \cite{PhysRevLett.116.033601}. A theoretical model based upon the polaron master equation (ME) formalism fully reproduces the experimental results and provides predictive power for the performance of a future optimised cavity-QD system.

\section{Methods}

\subsection{Theoretical Model}
\label{sec:Theory}
In this section we will outline the theoretical models used in the analysis of the coherence properties of the QD sample. 
We start by considering a two level system, with ground and single exciton states $\ket{0}$ and $\ket{X}$ respectively, and exciton energy $\hbar\omega_X$. The system is driven by a monochromatic continuous wave laser, with frequency $\omega_L$ and Rabi frequency $\Omega$, which in the dipole and rotating wave approximation can be described by the time-dependent system Hamiltonian~\cite{carmichael1999statistical}:
\begin{equation}
    H_\mathrm{S}(t) \approx \hbar\omega_X\sigma^\dagger\sigma + \frac{\hbar\Omega}{2}\left(\sigma e^{i\omega_\mathrm{L}t}+
    \sigma^\dagger e^{-i\omega_\mathrm{L}t}\right),
\end{equation}
where $\sigma=\ket{0}\bra{X}$ is the system dipole operator, and $\sigma_x = \sigma^\dagger + \sigma$.

The QD optical properties are strongly influenced by interactions with two environments: a  low-Q cavity mode, which induces strongly Purcell enhanced emission, and a phonon environment which describes the lattice vibrations of the surrounding material. 
In both cases, we can describe the environments as collection of bosonic modes, with Hamiltonian of the system and environment of the form:
\begin{eqnarray}
&H(t)=H_\mathrm{0}(t) + H_\mathrm{I}^\mathrm{EM} + H_\mathrm{I}^\mathrm{Ph},\\
&H_\mathrm{0}(t) = H_\mathrm{S}(t) +\sum_\mathbf{k}\hbar\nu_\mathbf{k} b^\dagger_\mathbf{k}b_\mathbf{k} + \sum_j \hbar\omega_j a_j^\dagger a_j,\\
&H_\mathrm{I}^\mathrm{EM} = \sum_j (f_j\sigma^\dagger a_j + f_j^\ast \sigma a_j^\dagger),\\
&H_\mathrm{I}^\mathrm{Ph}= \sigma^\dagger\sigma\sum_\mathbf{k} g_\mathbf{k}(b_\mathbf{k}^\dagger + b_{-\mathbf{k}})
+\sigma^\dagger\sigma\sum_\mathbf{k} \tilde{g}_{\mathbf{k},\mathbf{k}^\prime}(b_\mathbf{k}^\dagger + b_{-\mathbf{k}})
(b_{\mathbf{k}^\prime}^\dagger + b_{-\mathbf{k}^\prime}),
\end{eqnarray}
where we have introduced the bosonic annihillation operators $a_j$ and $b_\mathbf{k}$ associated with the normal modes of the electromagnetic and vibrational environments respectively.
The coupling to the optical environment is assumed to be of rotating-wave form, and is fully characterised by the spectral density $\mathcal{J}(\omega) = \sum_j \vert f_j\vert^2\delta(\omega-\omega_j)$, which for the low-$Q$ cavity studied here takes the form:
\begin{equation}
\mathcal{J}(\omega) = \frac{1}{\pi}\frac{2g^2\kappa}{(\omega - \omega_c)^2 + (\kappa/2)^2},
\label{eq:cavityJ}
\end{equation}
where $g$ is the light-matter coupling strength, $\kappa$ is cavity linewidth, and $\omega_c$ is its resonance.

The electron-phonon interaction, $H^\mathrm{Ph}_\mathrm{I}$, contains two contributions. The first is linear in phonon operators, and corresponds to real phonon processes~\cite{PhysRevLett.93.237401}, that is, the processes that involve the exchange of energy between the electronic states and the phonon environment. The strength of this interaction is determined by the matrix elements~\cite{mahan2000many} $g_\mathbf{k}=M_{e,\mathbf{k}}^{11}+M_{h,\mathbf{k}}^{11}$ for electrons ($e$) and holes ($h$), where for deformation potential coupling we have~\cite{Nazir_2016}:
\begin{equation}
    M_{a,\mathbf{k}}^{ij} = \sqrt{\frac{\nu_\mathbf{k}}{2\varrho c_s^2\mathcal{V}}}D_a\int  \psi_{ia}^\ast(\mathbf{r})
    \psi_{ja}(\mathbf{r})\mathrm{d^3}r,
\end{equation}
which is the matrix element corresponding to the phonon induced transition between the $i^\mathrm{th}$- and $j^\mathrm{th}$-electronic state.
Here, $\varrho$ is the mass density, $c_s$ is the speed of sound in the
material, and $\mathcal{V}$ is the phonon normalization volume. 
The matrix element depends on the wave function $\psi_{i,e/h}(\mathbf{r})$ of
the confined electron/hole and the corresponding deformation
potential $D_a$.


The second term, which is quadratic in phonon operators, describes virtual phonon transitions between the first exciton state ($s$-shell)
and higher lying excited states ($p$-shell) of the QD~\cite{PhysRevLett.93.237401}.
Intuitively, we may understand this term as a virtual scattering of a phonon with wavevector $\mathbf{k}$ into $\mathbf{k}^\prime$. 
This scattering process imparts a random phase kick
to the exciton, the cumulative effect of which is a temperature dependent broadening of the zero
phonon line~\cite{PhysRevLett.93.237401} and consequently a loss of photon coherence~\cite{PhysRevLett.118.233602,PhysRevLett.120.257401}.
This is governed by the effective coupling strength $\tilde{g}_{\mathbf{k},\mathbf{k}^\prime} = \sum_{a=e,h}\sum_{j>1}M_{a,\mathbf{k}}^{1j} M_{a,\mathbf{k}^\prime}^{j1}[\omega_{j}^a -\omega_{1}^a]^{-1}$, where $\omega_{j}^{e/h}$ is the energy of the $j^\mathrm{th}$-electron/hole state. For a detailed derivation and discussion of the quadratic coupling term, we refer the reader to Refs.~\cite{PhysRevLett.93.237401,PhysRevLett.118.233602,PhysRevLett.120.257401}.

It is important to note that while historically the linear electron-phonon coupling has been referred to as a pure-dephasing interaction~\cite{Axt05}, it does not lead to a temperature-dependent homogeneous broadening of the zero phonon line in the limit of weak driving~\cite{Mccutcheon13}.
For such processes, one must include the virtual phonon processes governed by the quadratic interaction.

\subsubsection{Polaron transformation and master equation}

In order to accurately describe the optical properties of a QD, we use the polaron framework~\cite{Iles-Smith2017}, where a unitary transformation $\mathcal{U} =\exp(\sigma^\dagger\sigma\otimes S)$, with $S= \sum_\mathbf{k}\nu_\mathbf{k}^{-1}g_\mathbf{k}(b^\dagger_\mathbf{k}-b_{-\mathbf{k}})$, is applied to the system-environment Hamiltonian~\cite{wurger98,Rae02,mccutcheon10}.
This
leads to a displaced representation of the phonon environment,
providing an optimized basis for a perturbative description of
the QD dynamics~\cite{mccutcheon10}. 
Importantly, this transformation naturally captures the non-Markovian relaxation behavior of the phonon
environment during exciton recombination~\cite{PhysRevB.95.201305,Iles-Smith2017,PhysRevLett.123.167403}.
In the polaron frame, we obtain the second-order master equation for the time evolution of the reduced state of the QD:
\begin{equation}\label{eq:master}
    \frac{\partial\rho(t)}{\partial t} = -i\left[
    \frac{\Omega_\mathrm{R}}{2}\sigma_x, \rho(t)\right]
     + \mathcal{K}[\rho(t)] + \frac{\gamma(T)}{2}\mathcal{L}_{\sigma^\dagger\sigma}[\rho(t)] + \frac{\Gamma}{2}\mathcal{L}_\sigma[\rho(t)],
\end{equation}
where $\mathcal{L}_O[\rho] = 2 O\rho O^\dagger -\{O^\dagger O, \rho\}$ is the Lindblad dissipator.
In Eq.~\ref{eq:master}, we have transformed the system into a rotating frame with respect to the laser frequency $\omega_\mathrm{L}$, which is assumed to be resonant with the polaron shifted transition frequency $\tilde\omega_\mathrm{X} = \omega_\mathrm{X}-\sum_\mathbf{k}\nu_\mathbf{k}^{-1}\vert g_\mathbf{k}\vert^2$.
The Rabi frequency, $\Omega_\mathrm{R} = \Omega B$, is renormalised by the Frank-Condon factor, which may be written as
\begin{equation}
    B = \exp(-\frac{1}{2}\int_0^\infty \mathrm{d}\nu~\frac{J(\nu)}{\nu^2}\coth(\frac{\nu}{2k_\mathrm{B}T})),
\end{equation}
where $T$ is the temperature and $k_\mathrm{B}$ Boltzmann's constant. 
Note we have taken the continuum limit of the phonon modes by introducing the phonon spectral density, $J(\nu) = \alpha \nu^3 \exp(-\nu^2/\nu_c^2)$, where $\alpha$ is the electron-phonon coupling strength and $\nu_c$ is the phonon cut-off frequency~\cite{Nazir_2016}.  

There are three dissipative mechanisms to consider in Eq.~\ref{eq:master}.
The second term in Eq.~\ref{eq:master}, is the polaron frame dissipator, $\mathcal{K}[\rho(t)] = -(\Omega/2)^2(\Gamma_0^{x}\left[\sigma_x, \sigma_x\rho(t)\right]+\left[\sigma_y, (\Gamma_s^y\sigma_z + \Gamma_c^y\sigma_y)\rho(t)\right] +\mathrm{h.c.})$, where the terms $\Gamma^a_0 = \int_0^\infty\Lambda_{aa}(\tau)d\tau$, $\Gamma^a_c = \int_0^\infty\Lambda_{aa}(\tau)\cos(\eta\tau)d\tau$, $\Gamma^a_s = \int_0^\infty \Lambda_{aa}(\tau)\sin(\eta\tau)d\tau$ may be understood as the rates at which transitions occur between the eigenstates of the system (i.e. the dressed states) induced by phonons~\cite{mccutcheon10}. 
These rates are set by the energy splitting of the system, and the correlation functions of the phonon environment in the polaron frame,
$
\Lambda_{xx}(\tau) =B^2(e^{\varphi(\tau)} + e^{-\varphi(\tau)} -2 )~~\textrm{and}~~\Lambda_{yy}(\tau) =B^2(e^{\varphi(\tau)} - e^{-\varphi(\tau)})
$, where $\varphi(\tau) = \int_0^\infty \nu^{-2}J(\nu)(\cos(\nu\tau) \coth(\nu/2k_\mathrm{B}T) - i\sin(\nu\tau))$.
The overall contribution of these phonon assisted transitions is scaled by the driving strength $\Omega^2$~\cite{mccutcheon10}.

The third term in Eq.~\ref{eq:master} gives the pure dephasing due to virtual phonon processes with rate~\cite{PhysRevLett.118.233602,Tighineanu18},
\begin{equation}
    \gamma(T) = \frac{\alpha\mu}{4\nu_c^4}\int_0^\infty \mathrm{d}{\nu}~\nu^{10}e^{-\nu^2/\nu_c^2}\left(\coth^2\left(\frac{\nu}{2 k_\mathrm{B}T}\right)-1\right),
    \label{eq:puredeph}
\end{equation}
where $\mu$ depends on the deformation potential coupling strength and spacing of the QD energy levels.
This dephasing rate is strongly temperature dependent and decays rapidly to zero for low temperatures. Physically this corresponds to an absence of phonons present to drive virtual transitions.

The final term in Eq.~\ref{eq:master} describes the optical emission through the cavity mode. Though in principle this emission rate $\Gamma$ will be temperature-dependent~\cite{Roy-Choudhury:15,Roy-ChouduryPRB}, for typical QD phonon parameters and for the cavity parameters of the current sample, this change is $\sim 3$~ps between 0-30~K, which is comparable to the uncertainty in the lifetime measurements.
We therefore neglect this effect, such that the spontaneous emission rate is $\Gamma\approx F_\mathrm{P}\Gamma_0$, where we have assumed the QD transition is on resonance with the cavity mode resulting in a Purcell factor  $F_\mathrm{P}=4g^2/\kappa$, and we have introduced the bulk emission rate $\Gamma_0$. 

\subsubsection{Coherent and incoherent scattering in the polaron frame} 
\label{sec:Theory_g1}
We are interested in understanding the impact that phonon coupling has on the optical properties of the QD, and specifically the coherence of scattered photons.
The coherence of a field is governed by the first order correlation function~\cite{carmichael1999statistical} $g^{(1)}(t,\tau) = \langle \hat{E}^\dagger(t+\tau)\hat{E}(t)\rangle$, where $\hat{E}(t)$ is the time-dependent field operator. 
This quantity contains all the information of the coherence of scattered photons and can be used, for example, to calculate the visibility of Hong Ou Mandel interference effects and thus photon indistinguishability~\cite{PhysRevA.69.032305,PhysRevB.95.201305,Iles-Smith2017}.
 Under CW driving consider here, we focus on the steady-state coherence $g^{(1)}(\tau) = \lim_{t\rightarrow\infty} g^{(1)}(t,\tau)$, which in the polaron frame can be divided into two contributions:
\begin{equation}
    g^{(1)}(\tau) = g^{(1)}_\mathrm{opt}(\tau) + g^{(1)}_\mathrm{SB}(\tau).
\end{equation}
The first contribution is associated to purely optical processes, and takes the form $g^{(1)}_\mathrm{opt}(\tau) = B^2\lim_{t\rightarrow\infty}\langle\sigma^\dagger(t+\tau)\sigma(t)\rangle$, which leads to the ZPL in the emission spectrum~\cite{PhysRevB.95.201305} and can be calculated using the quantum regression theorem~\cite{carmichael1999statistical}. 
Under CW driving, the optical contribution can be further sub-divided into a coherent and incoherent contribution: the coherent scattering is defined by the steady state $g^{(1)}_\mathrm{coh} = \lim_{\tau\rightarrow\infty}g^{(1)}_\mathrm{opt}(\tau)$, with the incoherent scattering naturally following as $g^{(1)}_\mathrm{inc}(\tau)=g^{(1)}_\mathrm{opt}(\tau)-g^{(1)}_\mathrm{coh}$.
In addition to direct optical scattering, there are also processes where a phonon is emitted or absorbed during the photon emission process, which leads to a broad spectral feature termed the phonon sideband~\cite{PhysRevB.95.201305}. 
This is captured by the 
$
    g^{(1)}_\mathrm{PSB}(\tau) = (\mathcal{G}(\tau)-B^2) g^{(1)}_\mathrm{opt}(\tau),
$
where $\mathcal{G}(\tau) = B^2\exp(\varphi(\tau))$ is the phonon correlation function.
The emission spectrum for each contribution can then be obtained by way of the Wiener Khinchin theorem~\cite{carmichael1999statistical}:
\begin{equation}
    S(\omega) = H(\omega) S_0(\omega) = H(\omega)\mathrm{Re}\left[\int_0^\infty g^{(1)}(\tau)e^{-i\omega\tau}~\mathrm{d}\tau\right],
    \label{eq:spectrum}
\end{equation}
where $H(\omega) = 8\pi g^2\kappa/[(\omega-(\omega_\mathrm{C} - \omega_\mathrm{X}))^2 + (\kappa/2)^2]$ is the cavity filter function~\cite{Iles-Smith2017}.

To compare with experiment, we are interested in the fractions of light emitted into the ZPL and through the coherent scattering channel~\cite{PhysRevLett.123.167403}. 
We therefore consider the partial powers, 
which are defined as the integral over the filtered spectrum associated with each emission channel, for example, the power through the PSB is given by $\mathcal{P}_\mathrm{PSB} = \int_{\infty}^\infty H(\omega)S_\mathrm{SB}(\omega)~\mathrm{d}\omega$, where $S_\mathrm{PSB}(\omega) = \mathrm{Re}[\int_0^\infty g^{(1)}_\mathrm{PSB}(\tau) e^{-i\omega\tau}~\mathrm{d}\tau]$. This allows us to define the filtered ZPL fraction as,
$    \mathcal{F}_\mathrm{ZPL} =\mathcal{P}_\mathrm{opt}/\mathcal{P}_\mathrm{Tot}$, where $\mathcal{P}_\mathrm{Tot} = \mathcal{P}_\mathrm{opt} + \mathcal{P}_\mathrm{PSB}$ is the total power emitted. In the absence of any spectral filtering from the cavity, the ZPL fraction reduces to the Frank-Condon factor $\mathcal{F}_\mathrm{ZPL}=B^2$.
To calculate the fraction of light emitted through coherent scattering processes we consider only photons emitted through the ZPL, such that $\mathcal{F}_\mathrm{coh} = \mathcal{P}_\mathrm{coh}/\mathcal{P}_\mathrm{opt}$, where $\mathcal{P}_\mathrm{coh} = \pi H(0)g^{(1)}_\mathrm{coh}$.



\subsection{Sample Characterisation}

\begin{figure}
    \centering
    \includegraphics{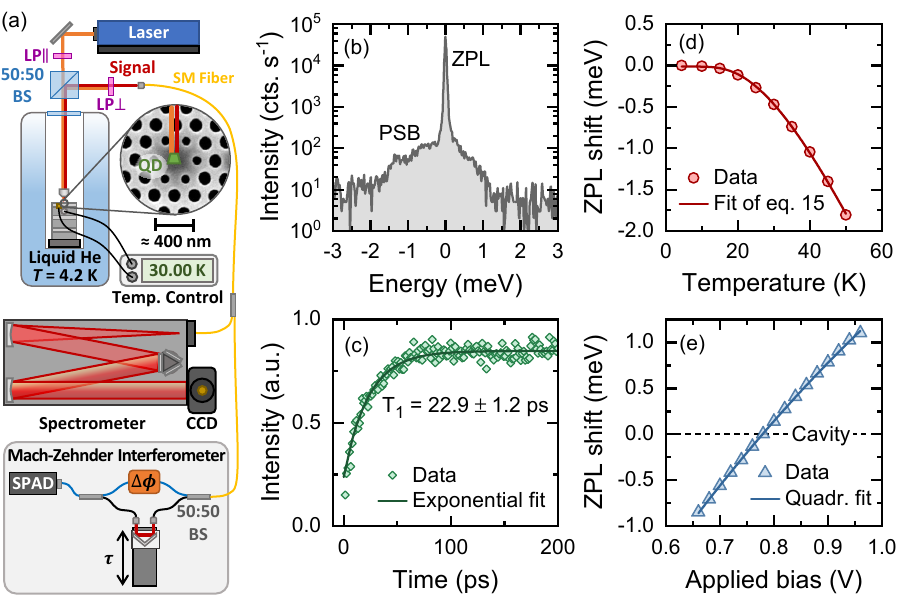}
    \caption{(a) Schematic of the experiment: BS - beam splitter, CCD - charge-coupled device (camera), LP - linear polarizer aligned either parallel ($\parallel$) or perpendicular ($\bot$) to input laser polarisation, 
	SM - single mode fiber, SPAD - single photon avalanche diode, $\Delta \phi$ - phase shift, $\tau$ - path length difference.
    (b) Experimental log-linear spectrum of the QD-cavity device under study at 4 K, showing the zero phonon line and phonon sideband. Narrow features at large detunings correspond to small detector background fluctuations that are only visible due to the logarithmic scale.
    (c) Pump-probe measurement of the cavity-enhanced QD radiative lifetime (green diamonds) fitted with an exponential decay (solid green line).
    (d) Measurement of the ZPL energy shift as a function of temperature (red circles) with a fit of a Bose-Einstein model according to eq. \ref{eq:redshift} (solid red line).
    (e) Measurement of the ZPL energy shift as a function of the bias voltage applied to the sample diode (blue triangles) with a quadratic fit (solid blue line).}
    \label{fig:setupandchar}
\end{figure}

Fig. \ref{fig:setupandchar}(a) shows a schematic of the experimental setup. The sample comprises of self-assembled InGaAs QDs embedded within a suspended 170 nm thick GaAs membrane. The membrane incorporates n- and p-doped GaAs layers, as well as AlGaAs tunnelling barriers, forming a p-i-n diode that can tune the QD emission by several meV using the quantum-confined Stark effect (QCSE). Using electron beam lithography and chemical etching nanofabrication techniques, H1 PhCCs are fabricated, consisting of a single point defect in a lattice of air holes (see inset in Fig. \ref{fig:setupandchar}(a)) . The device under study here comprises the neutral exciton state ($\ket{X}$) of a QD, weakly coupled to a resonant H1 PhCC (linewidth $2\hbar \kappa = 2.51~\mathrm{meV}$) that induces a significant Purcell enhancement. Further details of the sample and device under study may be found in Ref. \cite{Liu2018}.

The sample is located within a liquid helium bath cryostat at a base temperature of $T = 4.2~\mathrm{K}$. A feedback loop incorporating a resistive heater and a calibrated temperature sensor in the sample holder allows the temperature to be varied up to $50~\mathrm{K}$. The sample is excited by a tuneable single mode laser, with the emission separated from the laser by the use of orthogonal polarisers, producing a typical signal-to-background ratio of 100:1 for resonant excitation. The emission from the sample is then analysed either in the frequency domain with a grating spectrometer or in the time domain by a Mach-Zehnder interferometer that records the absolute value of the first-order correlation function ($\vert{g}^{(1)}(\tau)\vert$). Full details of the time domain measurement are presented in Section \ref{sec:ExperimentalMethod}.

Fig. \ref{fig:setupandchar}(b) shows a typical spectrum of the device under study with the heater switched off. The narrow ZPL and broad asymmetric PSB are both clearly visible when plotted on a logarithmic scale. To verify the Purcell enhanced lifetime of the QD transition under study, a pump-probe measurement is performed, plotting the ZPL intensity as a function of the separation of two resonant $\pi$-pulses in Fig. \ref{fig:setupandchar}(c) according to the method described in Ref. \cite{Liu2018}. An exponential fit to this data produces a value of $T_1 = 22.9 \pm 1.2~\mathrm{ps}$, in excellent agreement with the value of $22.7 \pm 0.9~\mathrm{ps}$ previously measured in Ref. \cite{Liu2018} that corresponds to a Purcell enhancement of $F_P = 43$.

To begin to investigate the behaviour of this device at elevated temperatures, the redshift of the ZPL is first characterised by fitting temperature-dependent spectra. The results are plotted in Fig. \ref{fig:setupandchar}(d) and show the characteristic non-linear behaviour where the redshift increases exponentially beyond an activation energy. The data agrees very well with a fit to a Bose-Einstein type model derived in Refs. \cite{PhysRevB.72.085328,Kroner2009}:
\begin{equation}
    \Delta(T) = - SE_{ph}\left(\mathrm{coth} \left( \frac{E_{ph}}{2k_b T}\right) -1 \right),
    \label{eq:redshift}
\end{equation}
where $S$ is a dimensionless coupling constant and the coth term describes the coupling of electrons to phonons of energy $E_{ph}$. The fitted values of $S = 0.6$ and $E_{ph} = 8.0~\mathrm{meV}$ are comparable to those found in previous studies of InGaAs QDs~\cite{Kroner2009}.

To independently study the influence of temperature on the emission properties of the cavity-QD system, it is necessary to compensate for the redshift of the QD with increasing $T$, such that the cavity remains resonant with the QD and maintains a constant Purcell enhancement. To achieve this, Fig.~\ref{fig:setupandchar}(e) shows a plot of the ZPL energy as a function of the bias voltage applied to the p-i-n diode. We observe a characteristic quadratic shift with voltage \cite{PhysRevLett.84.733} over a total range of around 2~meV. As the QD-cavity resonance condition lies close to the centre of this range at the base temperature, we are able to compensate over 1~meV of redshift by increasing the applied voltage as the temperature increases.

\subsection{Experimental Method}
\label{sec:ExperimentalMethod}

To investigate the coherence of the emitted photons as a function of temperature, we make a time-domain measurement of the first order correlation function $g^{(1)}(\tau)$ using a similar method to that described in Ref. \cite{PhysRevLett.123.167403}. This is performed using a Mach-Zehnder interferometer as shown in Fig. \ref{fig:setupandchar}(a). At each point in time ($\tau$), the phase between the two arms ($\Delta\phi$) is scanned, producing a set of interference fringes. The contrast ($v$) of these fringes is then evaluated according to
\begin{equation}
    v = \frac{I_{max}-I_{min}}{I_{max}+I_{min}},
    \label{eq:v}
\end{equation}
by using a generalised peak fitting routine to find the intensity at the local maximas ($I_{max}$) and minimas ($I_{min}$). By repeating this process for a range of $\tau$, the evolution of $v$ over the duration of the photon wavepacket can be plotted. The maximum resolvable contrast (defined as $1-\epsilon$) is limited by factors including imperfect mode overlap at the second beamsplitter, imperfect polarisation matching between the interferometer arms, and detector dark counts. As such, this varies depending upon experimental conditions but is around 0.95. The measured fringe visibility ($v$) as a function of $\tau$ can then be related to $g^{(1)}(\tau)$ by \cite{PhysRevLett.123.167403}:
\begin{equation}
    v(\tau)=(1-\epsilon)\frac{|g^{(1)}(\tau)|}{g^{(1)}(0)},
    \label{eq:vg1}
\end{equation}
demonstrating that once the interferometer imperfections are accounted for by the ($1-\epsilon$) term, $v(\tau)$ corresponds to the absolute value of the normalised coarse grained first-order correlation function. Significantly, unlike two-photon interference experiments which result in a single indistinguishability value \cite{PhysRevLett.116.033601,PhysRevB.97.195432,PhysRevLett.118.233602,PhysRevLett.118.253602}, this method allows the influence of real and virtual phonon transitions to be resolved independently according to their different characteristic timescales within $g^{(1)}(\tau)$.

\section{Results}
\label{sec:Results}

\begin{figure}
    \centering
    \includegraphics{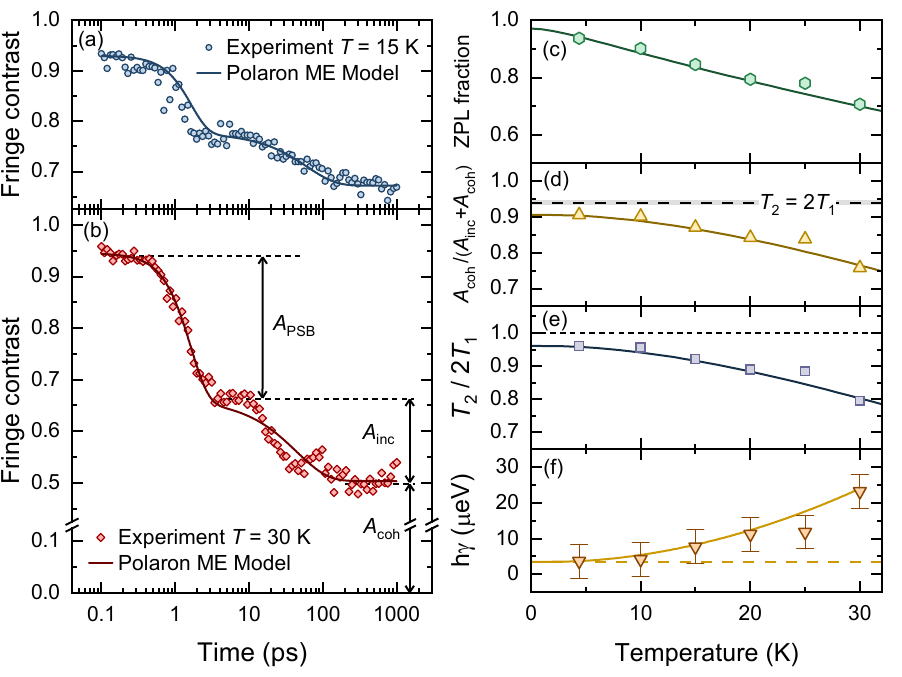}
    \caption{(a-b) Experimental fringe contrast measurement of the first order correlation function ($g^{(1)}(\tau)$) for temperatures of (a) 15 and (b) 30 K. Solid lines are from the Polaron model, using independently measured values aside from fitting to extract the phonon parameters $\alpha = 0.0446~\mathrm{ps^2}$, $\nu_c = 1.35~\mathrm{ps^{-1}}$ and $\mu = 0.005293~\mathrm{ps}^2$.
    (c-f) Coherence measures extracted from the temperature-dependent $g^{(1)}(\tau)$) measurements: 
    (c) ZPL fraction, (d) Coherent fraction, (e) $T_2/2T_1$ and (f) Pure dephasing rate $h\gamma$ as a function of temperature with results of the Polaron model (solid lines). Dashed lines in (d,e) indicate the ``ideal" values for $T_2=2T_1$ with the grey shading in (d) representing the uncertainty. The dashed line in (f) indicates the small additional non-thermal pure dephasing implied by the measurements. For all data without visible error bars, errors are comparable to the symbol size.}
    \label{fig:g1}
\end{figure}

Figs.~\ref{fig:g1}(a,b) show example measurements of fringe contrast as a function of time for temperatures of 15 K (a) and 30 K (b) respectively. The QD, laser and cavity are all mutually resonant, with this condition maintained as the temperature is increased by increasing the applied bias according to Figs.~\ref{fig:setupandchar}(d,e). The measurements are equivalent to a Fourier transform of the spectrum and exhibit 3 stage dynamics, a fast initial decay associated with the real PSB transitions \cite{PhysRevLett.123.167403}, an exponential decay with time constant $T_2$ corresponding to incoherent radiative decay of the ZPL, and a plateau at long timescales from coherent scattering. The coherently scattered photons inherit the coherence time of the laser \cite{Nguyen2011,PhysRevLett.108.093602}, which is sufficiently long to appear flat on this scale.
To extract the phonon parameters required by the polaron model, we fit the short-time dynamics of the $g^{(1)}$ function at $T=30~\mathrm{K}$ to the full correlation function derived in Sec.~\ref{sec:Theory_g1}. By focusing on times $\leq10~\mathrm{ps}$, we can consider only real phonon processes and neglect any virtual dephasing or optical decay. This allows us to do a two-parameter fit to the $g^{(1)}$ function, extracting an electron-phonon coupling strength $\alpha = 0.046~\mathrm{ps}^2$ and phonon cut-off frequency $\nu_c=1.35~\mathrm{ps}^{-1}$. These parameters agree closely with those independently extracted in a previous study on the same device \cite{PhysRevLett.123.167403} and are used for all other theoretical curves that follow.

By evaluating the mean values of the plateaus in the data ($100 - 500$ fs, $5 - 10$ ps and $200 - 1000$ ps), the amplitudes of each component ($A$) can be found as visualised by the arrows in Fig. \ref{fig:g1}(b). From this, the ZPL fraction can be found directly as
\begin{equation}
    \mathcal{F}_\mathrm{ZPL} = \frac{A_{inc}+A_{coh}}{A_{PSB}+A_{inc}+A_{coh}}.
    \label{eq:B2}
\end{equation}
Fig. \ref{fig:g1}(c) compares the theoretical predictions for $\mathcal{F}_\mathrm{ZPL}$ with the full experimental data-set of fringe contrast measurements from $4 - 30$~K. In this range, $\mathcal{F}_\mathrm{ZPL}$ varies in an almost linear manner \cite{Denning:20}, reducing from $0.94 \pm 0.01$ at 4 K to $0.71 \pm 0.01$ at 30 K due to the increasing probability of the real phonon transitions at elevated temperatures.

Whilst the ZPL fraction is invariant with excitation conditions \cite{PhysRevLett.123.167403,PhysRevB.95.201305,PhysRevLett.123.167402}, the coherent fraction is very sensitive to both the driving strength (the phonon renormalised Rabi frequency - $\Omega_\mathrm{R}$) and the emitter coherence \cite{CohenTannoudhiAtom}:

\begin{equation}
    \mathcal{F}_{coh} = \frac{A_{coh}}{A_{inc}+A_{coh}} = \frac{T_2}{2T_1}\frac{1}{1+\Omega^2_\mathrm{R} T_1 T_2}.
    \label{eq:cohFrac}
\end{equation}
\noindent
Therefore, Eq. \ref{eq:cohFrac} illustrates that by maintaining constant values of $\Omega_\mathrm{R}$ and $T_1$, the coherently scattered fraction can be a sensitive probe of the QD coherence time $T_2$. $T_1$ is kept constant at the value of 22.9 ps measured in Fig. \ref{fig:setupandchar}(c) by the aforementioned technique of balancing the QD redshift with temperature (Fig. \ref{fig:setupandchar}(d)) with an equivalent blueshift from an increased applied bias (Fig. \ref{fig:setupandchar}(e)), keeping the QD resonant with the laser and cavity. Meanwhile, the Rabi frequency is calibrated at the beginning of each measurement by recording a series of Mollow triplet \cite{ulhaq2012cascaded} spectra at different excitation powers. Plotting half of the Mollow side-peak splitting (equal to $\Omega_\mathrm{R}$) vs. the square root of the laser power ($P^{1/2}$) allows for a linear fit linking laser power to Rabi frequency. To give high sensitivity through a large coherent fraction, a Rabi energy of $\hbar \Omega_\mathrm{R} = 5.11~\upmu\mathrm{eV}$ is used throughout.

Applying this approach, Fig. \ref{fig:g1}(d) shows the coherent fraction as a function of temperature, evaluated according to Eq.~\ref{eq:cohFrac}. The dashed horizontal line corresponds to the theoretical maximum coherent fraction of $0.940 \pm 0.007$, evaluated from the RHS of Eq.~\ref{eq:cohFrac} by taking $T_2 = 2T_1$. The experimental values begin at $0.906 \pm 0.014$ at 4 K, falling to $0.758 \pm 0.016$ at 30 K as dephasing of the ZPL becomes more significant. Whilst the value at 4 K is not quite transform-limited, we note that our measurement technique is a particularly stringent test of coherence as it is sensitive to any dephasing within the experiment duration (seconds). 
Most previous studies have used two photon interference methods that exclude any processes on timescales greater than the nanosecond separation between subsequent photons \cite{PhysRevB.72.085328,PhysRevB.97.195432,PhysRevLett.116.033601,PhysRevLett.118.253602}. When the timescale is extended in such measurements, a small decay in visibility is often observed \cite{PhysRevLett.116.213601}, including in previous two photon interference measurements on this sample \cite{Liu2018}. This effect likely originates from charge or spin noise \cite{Kuhlmann2013}, phenomena which may also explain the small non-thermal dephasing observed at low temperatures here.

With the measurement of coherent fraction, it is now possible to rearrange Eq. \ref{eq:cohFrac} to find
\begin{equation}
    \frac{T_2}{2T_1} = \frac{\mathcal{F}_{coh}}{1 - 2\Omega_\mathrm{R}^2 T_1^2 \mathcal{F}_{coh}}.
    \label{eq:T2coh}
\end{equation}
Using this equation with the previously found values of $T_1$, $\mathcal{F}_{coh}$ and $\Omega_\mathrm{R}$, Fig. \ref{fig:g1}(e) shows $T_2/2T_1$ as a function of temperature. At 4 K, $T_2/2T_1 = 0.961 \pm 0.014$, decreasing to $T_2/2T_1 = 0.796 \pm 0.018$ by 30 K. It is also then possible to extract the pure dephasing rate $\gamma = 1/T_2^*$ from Eq. \ref{eq:T2}, with the results plotted in Fig. \ref{fig:g1}(f). To find the prefactor $\mu$ for the virtual phonon dephasing described by Eq. \ref{eq:puredeph}, we fit to this experimental data, adding an additional constant value ($3.5 ~\upmu\mathrm{eV}$ - dashed line in Fig. \ref{fig:g1}(f)) to describe the small non-thermal dephasing implied by Fig. \ref{fig:g1}(d/e). The fitted value is $\mu=0.00529~\mathrm{ps}^2$.

\section{Discussion}

\begin{figure}
    \centering
    \includegraphics{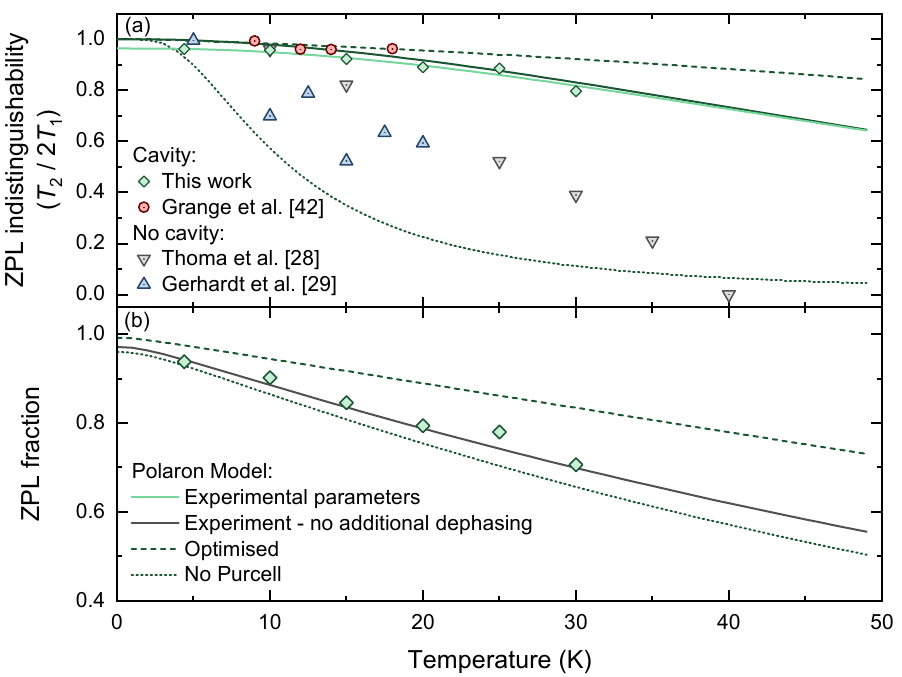}
    \caption{Comparison of (a) ZPL indistinguishability ($T_2/2T_1$) and (b) ZPL fraction from this study (green diamonds) with prior work (red circles, blue triangles and grey inverted triangles) and the Polaron model (lines). For the results from prior studies, $T_2/2T_1$ values are equated to two photon interference visibilities measured through a narrow spectral filter that removes the PSB. The polaron model as shown in Fig. \ref{fig:g1}(c-f) is the light green solid line, whilst the dark green solid line is the same parameters but with the additional non-thermal dephasing removed for comparison. The dark green dotted line shows the results of the Polaron model with the same parameters but without any Purcell enhancement. The dark green dashed line uses the same phonon parameters but reduces $\kappa$ to increase the Purcell factor to $F_P = 200$, chosen to be just below the onset of strong QD-cavity coupling.}
    \label{fig:comparison}
\end{figure}

In the results section, the temperature dependence of the ZPL fraction and ZPL coherence ($T_2/2T_1$) were measured in the range of 4 - 30 K. In this range, it was found that the ZPL fraction decayed almost linearly from 0.937 to 0.7, whilst $T_2/2T_1$ decreases from 0.961 to 0.798 with the gradient increasing at higher $T$. Recalling Eq. \ref{eq:V}, we note that $T_2/2T_1$ is equivalent to the indistinguishability of photons emitted through the ZPL, as would be measured in a two photon interference experiment with a spectral filter that removes the PSB component. Using this fact, Fig. \ref{fig:comparison}(a) presents a comparison of ZPL indistinguishability between our measurements of $T_2/2T_1$ vs. $T$ and equivalent previous two photon interference experiments \footnote{We note that as previously discussed in section \ref{sec:Results}, such two photon interference measurements are not sensitive to dephasing processes that can occur on longer timescales. As such, for comparison in Fig. \ref{fig:comparison} we have plotted our polaron model both with (light green solid line) and without (dark green solid line) the small non-thermal pure dephasing that was inferred in Figs. \ref{fig:g1}(d-f).}. Thoma et al. \cite{PhysRevLett.116.033601} and Gerhardt et al. \cite{PhysRevB.97.195432} (grey triangles and blue inverted triangles respectively in Fig. \ref{fig:comparison}(a)) consider QDs without any significant Purcell enhancement, therefore it is unsurprising that their values for $T_2/2T_1$ rapidly fall away from those measured here (green diamonds) as $T$ increases. For comparison, at $T = 30~\mathrm{K}$, Thoma et al. \cite{PhysRevLett.116.033601} measure $T_2/2T_1 = 0.39$, half the value measured here.

Meanwhile, Grange et al. \cite{PhysRevLett.118.253602} (red circles) measured values ranging between 0.99 and 0.96 in the range 9 - $18~\mathrm{K}$ for a QD micropillar system with a Purcell factor of 20, compared to a Purcell factor of 43 and temperature range of 4 - $30~\mathrm{K}$ for the device studied here. Whilst direct comparisons are difficult due to the much smaller temperature range, the local gradient in $T_2/2T_1$ appears lower despite the lower Purcell factor, suggesting that the underlying thermal ZPL broadening of the sample used in Ref. \cite{PhysRevLett.118.253602} may be lower than the QD studied here. Considering other prior studies on different QD samples, we also note that the dephasing value of $11.8~\upmu\mathrm{eV}$ at $T = 25~\mathrm{K}$ in Fig. \ref{fig:g1}(f) is significantly larger than the $\sim 4~\upmu\mathrm{eV}$ measured in a previous four-wave mixing study at the same temperature \cite{PhysRevLett.93.237401}, whilst our extracted value of $\mu$ is an order of magnitude larger than that found from two photon interference experiments in Ref. \cite{PhysRevLett.118.233602}. These comparisons suggest significant variations in the thermal dephasing rates of different QDs. A possible explanation lies in the theoretical thermal dephasing rate $\gamma(T)$ given by Eq. \ref{eq:puredeph}, this expression contains both the cut-off frequency $\nu_c$, and the prefactor $\mu$ that varies with the QD energy level spacing. As both of these quantities depend upon the QD size and shape, significant variation in the thermal dephasing could be explained by variations in QD geometry between different samples.
Whilst detailed consideration of QD structure is beyond the scope of this work, it may provide an interesting direction for further study.

Exploiting the excellent agreement between the polaron model (solid green lines in Fig. \ref{fig:comparison}) and experimental results, we now model two additional scenarios; an optimised QD-cavity device with $F_P$ increased to 200 by reducing $\kappa$ and a bare QD without any Purcell enhancement. These models use the QD and phonon parameters found from fitting the experimental data, varying only the cavity parameters and setting any non-thermal dephasing to zero. Considering first the case without Purcell enhancement (dotted green lines in \ref{fig:comparison}), we note that in Fig. \ref{fig:comparison}(a) the ZPL coherence falls rapidly, reaching $T_2/2T_1 = 0.11$ at $T = 30~\mathrm{K}$. This illustrates the importance of the Purcell enhancement - our QD-cavity device improves on this value by more than a factor of 7. In addition, it is noticeable that without Purcell enhancement, $T_2/2T_1$ falls much faster with increasing temperature than the cavity-free measurements of Refs. \cite{PhysRevLett.116.033601,PhysRevB.97.195432}, providing further evidence that the underlying thermal dephasing rate of this QD appears significantly greater than previous studies.

Meanwhile, the dashed line in Fig.~\ref{fig:comparison} shows the same model but for an optimised cavity with $F_P = 200$ by reducing $\kappa$. This value is chosen to maximise the Purcell factor whilst ensuring that the cavity-QD system does not enter the strong coupling regime where photon coherence begins to decrease again \cite{Iles-Smith2017}. For these parameters, the increased Purcell enhancement significantly improves $T_2/2T_1$ from 0.83 to 0.92 at $T = 30~\mathrm{K}$ when compared to the model for the sample cavity parameters. The magnitude of this difference continues to increase with temperature. When considering the fraction of light emitted into the ZPL (Fig.~\ref{fig:comparison}(b)), a small difference ($\sim 0.04$ at 30 K) is observed between the sample parameters and the "no Purcell" model. This is due to the photonic spectral density of the cavity (Eq.~\ref{eq:cavityJ}) removing some of the PSB contribution according to Eq. \ref{eq:spectrum}. The effect is relatively small as the half-width of the cavity ($\kappa$) is comparable to the phonon cut-off frequency $\nu_c$. For the optimised system with reduced $\kappa$, the ZPL fraction at 30 K increases significantly from 0.70 to 0.83 due to the five-fold reduction in cavity linewidth. Whilst it seems intuitive that further reducing $\kappa$ will continue to be advantageous in this way, the onset of strong QD-cavity coupling ultimately degrades the photon coherence, leading to a fundamental trade-off between indistinguishability and efficiency \cite{Iles-Smith2017}. Unlike Fig.~\ref{fig:comparison}(a), it is not possible to easily compare ZPL fraction with previous studies as two photon interference measurements cannot easily isolate the PSB contribution.

\section{Conclusion}

In conclusion, we have demonstrated a QD-nanocavity device that exploits a large Purcell enhancement to achieve a high degree of photon coherence at elevated temperatures. Our novel experimental approach based upon time-domain measurement first-order correlation function is able to distinguish between contributions from real and virtual phonon-mediated transitions in a single measurement. 
Exploiting this, at a temperature of $30~\mathrm{K}$ that is compatible with the operational temperature of compact cryocoolers, we measure a ZPL coherence of $T_2/2T_1 = 0.80$ with a ZPL fraction of 0.71, compared to the $T_2/2T_1 = 0.11$ predicted by our model in absence of the Purcell enhancement. These experimental results are achieved despite the studied QD device exhibiting significantly stronger thermal dephasing than was observed in previous QD studies, a result that indicates that the QD size/shape may play a role in determining the magnitude of phonon dephasing. We have also developed a theoretical model based upon the polaron framework that fully captures the temperature-dependent phonon processes in order to reproduce our experimental results. The excellent agreement between theory and experiment provides predictive power, allowing us to simulate an optimised cavity-QD device that can achieve $T_2/2T_1 = 0.92$ with a ZPL fraction of 0.83, while fully accounting for electron-phonon processes using experimentally measured parameters.

Whilst indistinguishability requirements are application specific, experiments have successfully demonstrated the quantum interference phenomenon of boson sampling with a QD source exhibiting indistinguishabilities in the range $0.5 - 0.7$~\cite{PhysRevLett.118.130503}, suggesting that even our current device could perform such experiments at $30~\mathrm{K}$ when combined with a spectral filter to remove some of the PSB. We believe that the theoretical and experimental methods developed here can support the development of a new generation of cavity-QD quantum light sources, meeting both the photon coherence and SWAP requirements of emerging optical quantum technologies. Furthermore, with some adaptations to the specifics of phonon interactions in different materials, our methods can readily be applied to other emerging solid-state quantum emitter systems in materials such as diamond \cite{benedikter2017cavity}, silicon \cite{redjem2023all} and 2D materials \cite{doi:10.1021/acs.nanolett.2c03151}.

\section*{Acknowledgements}

The authors acknowledge Edmund Clarke and Ben Royal respectively for growth and nanofabrication of the sample as previously reported in Ref. \cite{Liu2018}. The authors also thank Catherine Philips for contributions to developing the time-domain measurement technique previously reported in Ref. \cite{PhysRevLett.123.167403}, and Mark Fox,
Maksym Sich and Scott Dufferwiel for interesting discussions regarding the potential operation of QDs in compact cryocoolers. A.J.B. gratefully acknowledges the support of the EPSRC (UK) through the Quantum Technology Fellowship EP/W027909/1 and Programme Grant EP/N031776/1, in addition to support from Research England through the National Productivity Investment Fund. 

\section*{References}
\bibliographystyle{iopart-num.bst}
\bibliography{bibliography}

\end{document}